\documentclass[journal,10pt]{IEEEtran}

\usepackage{setspace}

\usepackage{graphics,
           psfrag,
           epsfig,
           amsthm,
           cite,
           amssymb,
           url,
           dsfont,
           subfigure,
           balance,
           enumerate,
           color,
           setspace
}
\usepackage{amsmath}

\usepackage{epstopdf}

\newtheorem{lemma}{Lemma}

\newtheorem{theorem}{Theorem}

\newtheorem{proposition}{Proposition}

\newtheorem{remark}{Remark}

\newcommand{\eref}[1]{(\ref{#1})}

\newcommand{\pref}[1]{Proposition~\ref{#1}}
\newcommand{\cref}[1]{Constraint~\ref{#1}}

\newcommand{\lref}[1]{Lemma~\ref{#1}}

\newcommand{\ignore}[1]{}

\IEEEoverridecommandlockouts
\usepackage{algorithm}
\usepackage{algpseudocode}
  
\algnewcommand\algorithmicforeach{\textbf{Until :}}
\algnewcommand\algorithmicendif{\textbf{End}}
\algnewcommand\ForEach{\item[ \algorithmicforeach]}
\algnewcommand\EndiFF{\item[ \algorithmicendif]}
\usepackage{mathtools}

\allowdisplaybreaks
\usepackage{flushend}
\usepackage{lipsum}

\title{\LARGE \bf 
Secret-Key Agreement with Public Discussion subject to an Amplitude Constraint
 }
\author{Marwen Zorgui,~\IEEEmembership{Student Member,~IEEE}, 
Zouheir Rezki,~\IEEEmembership{Member,~IEEE}, 
Basel Alomair,~\IEEEmembership{Member,~IEEE}
and Mohamed-Slim Alouini,~\IEEEmembership{Fellow,~IEEE}
\vspace{-0.7cm}
}

\begin{document}
\maketitle
\begin{abstract}
This paper considers the problem of secret-key agreement with public discussion subject to a peak power constraint $A$ on the channel input. The optimal input distribution is proved to be discrete with finite support. The result is obtained by first transforming the secret-key channel model into an equivalent Gaussian wiretap channel with better noise statistics at the legitimate receiver and then using the fact that the optimal distribution of the Gaussian wiretap channel is discrete. To overcome the computationally heavy search for the optimal discrete distribution, several suboptimal schemes are proposed and shown numerically to perform close to the capacity. Moreover, lower and upper bounds for the secret-key capacity are provided and used to prove that the secret-key capacity converges for asymptotic high values of $A$, to the secret-key capacity with an average power constraint $A^2$.
Finally, when the amplitude constraint A is small ($A \to 0$), the secret-key capacity is proved to be asymptotically equal to the capacity of the legitimate user with an amplitude constraint A and no secrecy constraint.
\end{abstract}
\section{Introduction}
\label{S1}
This work considers the channel-type model with wiretapper for secret-key agreement (CW model) \cite{common_randomness,wongsecret2009} which consists of a transmitter, a legitimate receiver and an eavesdropper. An authenticated and noise-free public channel is made available to the legitimate users, over which they can exchange an arbitrary number of public messages that are assumed to be perfectly observed by the eavesdropper. In this model, the objective of the legitimate users is to agree on a sequence of bits that should be kept secure from the eavesdropper. 
In \cite{common_randomness}, the authors established the single-letter capacity expression of the CW model for discrete channel input. Under continuous channel input and an average power constraint, the secret-key capacity is shown to be achieved by a Gaussian input with full power \cite{wongsecret2009}.
In \cite{Maurer}, the achievability of a positive key rate was proved for memoryless binary channels when the destination and the eavesdropper channels are conditionally independent.
In this work, the capacity of the CW model subject to a peak amplitude constraint on the input is studied. 
The motivation behind this work is twofold. First, determining the capacity of a communication channel subject to various input constraints is a classical problem of information theory. Second, the need for incorporating an amplitude constraint is dictated by the fact that several practical communication systems are subject to a peak-power constraint or/and an average power constraint. For instance, in visible light communications, and for safety reasons, the modulating signal must satisfy an amplitude constraint, rather than a conventional average power constraint.

The problem of finding capacity-achieving distributions with an amplitude constraint has been addressed for several channel models. In \cite{smith1971information}, Smith studied the capacity of the scalar Gaussian channel subject to both a peak power constraint and an average power constraint. The capacity-achieving distribution, rather surprisingly, was shown to be discrete, with a finite number of probability mass points. Following the same approach, 
several works such in \cite{chan2005capacity,discreteness_study,perf_analysis,
agrawal2011noncoherent} showed that the optimal capacity-achieving distributions are also discrete for various channel models.
In \cite{ozel2015gaussian}, the authors considered the Gaussian wiretap channel with an amplitude constraint on the channel input. The entire rate-equivocation region of such channel was shown to be achieved by discrete input distribution with finite support. 

In this paper, the capacity-achieving input distribution for the Gaussian CW model is proved to be also discrete with finite support. The result is obtained by first transforming the secret-key channel model into an equivalent Gaussian wiretap channel with better noise statistics at the legitimate receiver and then using the fact that the optimal distribution of the Gaussian wiretap channel is discrete.
The capacity-achieving probability distribution has to be computed numerically relying on a necessary and sufficient Karush-Kuhn-Tucker (KKT) optimality condition. To circumvent the numerical complexity associated with finding the capacity-achieving input distribution, suboptimal schemes are proposed and shown numerically to perform close to the capacity. 
Finally, lower and upper bounds for the secret-key capacity are derived. Based on these bounds, the behavior of the secret-key capacity for asymptotic high and low values of $A$ is characterized. For instance, the secret-key capacity with an amplitude constraint converges for asymptotic high values of $A$, to the secret-key capacity with an average power constraint $A^2$. When the amplitude constraint A is small ($A \to 0$), the secret-key capacity is asymptotically equal to the capacity of the legitimate user with an amplitude constraint A and no secrecy constraint.
\section{System model}\label{S2}
The Gaussian CW model for secret-key agreement \cite{common_randomness} is defined by

\vspace{-.4cm}
\begin{small}
\begin{equation}
\label{system_model}
\begin{split}
Y_i&=X_i+ N_{D_i}, \qquad i=1 \ldots n\\
Z_i&=X_i+ N_{E_i}, \qquad i=1 \ldots n,
 \end{split}
\end{equation}
\end{small}

\vspace{-.4cm}
where $n$ is the number of channel uses. $X_i$, $Y_i$ and $Z_i$ denote the channel input, the legitimate receiver's observation, and the eavesdropper's observation, respectively. $N_{D_i}$ and $N_{E_i}$ are independent and identically distributed zero-mean Gaussian random variables with variances $\sigma_D^2$ and $\sigma_E^2$, respectively. 
The channel input $X_i$ is assumed to have an amplitude constraint

\vspace{-.4cm}
\begin{small}
\begin{align}
\label{peak_constraint}
\lvert X_i \lvert \le A,\qquad i=1,\ldots n. 
\end{align}
\end{small}

\vspace{-.4cm}
A single-letter characterization of the secret-key capacity of the CW model with conditional probability distribution function (PDF) $ p(y,z|x)=p(y |x) p( z|x) $ is established in \cite{common_randomness} for the case of discrete channel alphabets. Following the extension to continuous channel alphabets with input average power constraint\cite[Theorem 1]{wongsecret2009}, the secret-key capacity with a peak amplitude constraint can be expressed in a single-letter form as follows. 
\begin{lemma}
\normalfont
The secret-key capacity of the CW model $(X,Y,Z)$ with conditional PDF $p(y,z|x)=p(y |x) p( z|x)$ and a peak amplitude constraint is given by

\vspace{-.4cm}
\begin{small}
\begin{align}
\label{secret_key_capacity}
C_k= \underset{ F_X \in \Omega}{\sup} [ \mathbb{I} (X;Y)- \mathbb{I} (Y;Z) ]= 
\underset{ F_X \in \Omega}{\sup}   \mathbb{I} (X;Y|Z) 
,
\end{align}
\end{small}

\vspace{-.4cm}
where the supremum is over all input probability distributions $F_X$ in 
{\small $\Omega \triangleq \left\{ F_X: \int^A_{-A} dF_X(x)=1 \right\}  $}.
\end{lemma}
The secret-key capacity can be further expressed as 

\vspace{-.4cm}
\begin{small}
\begin{align}
\label{secret_key_expression}
C_k= \underset{ F_X \in \Omega}{\sup} [ \mathbb{I}(X;Y,Z)-\mathbb{I}(X;Z)].
\end{align}
\end{small}

\vspace{-.4cm}
In the sequel, the input distribution achieving the secret-key capacity is characterized. 
\section{Secret-key capacity}\label{S3}
The capacity-achieving probability distribution is discrete with a finite number of probability mass points. 
This is formalized in the following theorem. 
\begin{theorem}
\label{discreteness_theorem}
\normalfont
Let $F_0$ be a solution to \eref{secret_key_capacity}. Then, the support set of $F_0$, $\mathcal{S}_{F_0}$, is a finite set of points.
\end{theorem}
\begin{IEEEproof}
From \eref{secret_key_expression}, the secret-key rate $\mathbb{I}(X;Y,Z)-\mathbb{I}(X,Z)$ is equivalent to a secrecy rate in which the destination is provided with the eavesdropper observation. In particular, $\mathbb{I}(X;Y,Z)$ is the rate of channel given by $Y_{DE}= h X + N$, such that $Y_{DE}=\begin{pmatrix}
 Y \\ 
 Z
 \end{pmatrix}, h=  \begin{pmatrix}
 1 \\ 
 1
 \end{pmatrix}$, and $N= \begin{pmatrix}
 N_D \\ 
 N_E
 \end{pmatrix}$. Note that that covariance of N is given by $\Delta= \begin{pmatrix}
 \sigma_D^2 & 0 \\ 
  0 & \sigma_E^2
 \end{pmatrix}$. Let $B=\begin{pmatrix}
 \frac{1}{\sigma_D} & 0 \\ 
  0 & \frac{1}{\sigma_E}
 \end{pmatrix}$. It follows that $B \Delta B^T= I$. As $Bh$ is a column vector, its singular-value decomposition is given by $B h= V \lVert Bh \rVert 1$, with $V=\frac{Bh}{\lVert Bh \rVert}$. Therefore, 
 
 \vspace{-.4cm}
\begin{small}
 \begin{align}
\label{equivalent_main_channel}
\mathbb{I}(X;Y,Z)&=\mathbb{I}(X;h X+N)=\mathbb{I}(X;V^T B h X+V^T B N) \nonumber\\
&=\mathbb{I}(X; X+ \underbrace{\frac{1}{ \lVert B h \rVert} V^T B N}_{N_{eq}}). 
\end{align}
 \end{small} 
 
 \vspace{-.4cm}
Therefore, the rate $\mathbb{I}(X;Y,Z)$ corresponds equivalently to a channel with input $X$ and a single output, $Y_{eq}= X+ N_{eq}$ and with an equivalent noise variance given by

\vspace{-.4cm}
\begin{small}
\begin{align}
\label{equivalent_noise}
\mathrm{var} (N_{eq})=\frac{1}{\lVert B h \rVert^2} V^T B \Delta B^T V= \frac{1}{\frac{1}{\sigma_D^2}+\frac{1}{\sigma_E^2}} \triangleq \sigma_{DE}^2 .
\end{align}
\end{small}

\vspace{-.4cm}
This says that the secret-key model is equivalent to a Gaussian wiretap channel with the legitimate receiver having a better noise variance given by \eref{equivalent_noise}. Since $\mathrm{var}(N_{eq}) < \sigma_E^2$, the eavesdropper's channel is a stochastically degraded channel. Therefore, the optimization problem is equivalent to a Gaussian wiretap channel with an amplitude constraint problem, for which the existence of a capacity-achieving distribution and its discreteness follow along similar lines as in \cite[Proof of Corollary 1]{ozel2015gaussian}.
\end{IEEEproof}

\begin{remark}
\normalfont
The equivalent channel given by \eref{equivalent_main_channel} may also be established using the fact that a sufficient statistic to decode $X$ reliably from $(Y,Z)$ is $\frac{Y}{\sigma_D^2}+\frac{Z}{\sigma_E^2}$, along with the fact that a sufficient statistic preserves the mutual information. It follows

\vspace{-.4cm}
\begin{small}
\begin{align}
\mathbb{I}(X;Y,Z)=\mathbb{I}(X;\frac{Y}{\sigma_D^2}+\frac{Z}{\sigma_E^2})=\mathbb{I}(X;X+N_{eq}).
\end{align}
\end{small}

\vspace{-.4cm}
\end{remark}
The secret-key capacity-achieving distribution can be found
via a classical search algorithm as in \cite{smith1971information}. Briefly, at each step, the number of mass points is increased by unity and the optimal input distribution for given the current number of mass points is determined numerically. The process is repeated until the KKT conditions given by \cite[Equations 20 and 21]{ozel2015gaussian} are satisfied. However, such an optimization is computationally hungry, especially as the value of $A$ increases. This motivates the search for suboptimal, but efficient schemes that circumvent the complexity of finding the capacity-achieving distribution. 
\section{Suboptimal Input Distributions and bounds}\label{S4}
For each of the following schemes, the secret-key rate is evaluated using a unified approach. Following the proof of Theorem 1, the secret-key channel is transformed into an equivalent Gaussian wiretap channel. Starting from \eref{secret_key_expression} and using \eref{equivalent_noise}, the secret-key rate can be expressed as

\vspace{-0.4cm}
\begin{small}
 \begin{align}
 \label{equivalent_wiretap}
\mathbb{I}(X;Y,Z)-\mathbb{I}(X,Z)&=\mathbb{I}(X,X+N_{eq})-\mathbb{I}(X;X+N_{E}) \\
&=h(X+N_{eq})-h(X+N_{E})+\frac{1}{2} \log(\frac{ \sigma_E^2}{\sigma_{DE}^2}).
\nonumber
\end{align}
\end{small}

\vspace{-0.4cm}
Thus, in order to evaluate the secret-key rate, one only needs to evaluate the PDF of a generic random variable $Y=X+N$ where $N$ is an independent Gaussian noise and $X$ follows a specific distribution that satisfies the peak constraint \eref{peak_constraint}.
\subsection{Suboptimal input distributions}
One may consider suboptimal discrete input distributions such as maxentropic distributions considered in 
\cite{max_entropic}. Maxentropic distributions are probability distributions over a fixed number $K$ of equally spaced mass points that maximize the entropy of all admissible discrete distributions. In the current setting, maxentropic distributions are simply uniform distributions over the equally spaced mass points. One can then optimize over the number of mass points $K$ to achieve better performance. 

The continuous uniform distribution maximizes the entropy of the input subject to an amplitude constraint \cite{conrad2013probability}.
In order to evaluate the secret-key achieved by a uniform input distribution, one needs to evaluate the PDF of a random variable $T=X+N$ where $N$ is an independent Gaussian noise. The following lemma characterizes such PDF.
\begin{lemma}
\normalfont
\label{uniform_pdf}
Let $X  \sim \mathcal{U}(-A,A)$, $N \sim \mathcal{N}(0,\sigma^2)$ and $T=X+N$. Then the PDF of $T$, $p_T(t)$, is given by

\vspace{-.4cm}
\begin{small}
\begin{align}
p_T(t)=\frac{1}{2A} \left[ \mathcal{Q}(\frac{-A-t}{\sigma})-\mathcal{Q}(\frac{A-t}{\sigma}) \right],
\end{align}
\end{small}

\vspace{-.4cm}
where $\mathcal{Q}$ denotes the Q-function defined as $\mathcal{Q}(x)=\frac{1}{\sqrt{2 \pi}} \int_{x}^{+\infty}\exp(-\frac{u^2}{2}) du$ and $\mathcal{U}(-A,A)$ designates the uniform PDF over $[-A,A]$.
\end{lemma}
In case of an average power constraint only, the optimal input distribution is Gaussian. This suggests considering a truncated Gaussian input distributions, denoted as $\mathcal{N}_A(0,\sigma_x^2)$. The PDF of the truncated Gaussian is given as

\vspace{-.4cm}
\begin{small}
\begin{align}
p_X(x)= \frac{1}{\sigma_x D}\  \phi(\frac{x}{\sigma_x})\  \forall x \in [-A, A],
\end{align}
\end{small}

\vspace{-.4cm}
where $D= \Phi(\frac{A}{\sigma_x})- \Phi(-\frac{A}{\sigma_x})$, $\phi(\cdot)$ and $\Phi(\cdot)$  correspond to the PDF and the cumulative distribution function of the standard normal distribution, respectively. 
Unlike the uniform distribution, such an input distribution provides the flexibility of optimizing over its parameter $\sigma_x^2$ to achieve higher rates \cite{truncated_gauss}. As in the previous section, one needs to evaluate the PDF of a generic random variable $T=X+N$ where $X$ is a truncated Gaussian and $N$ is an independent Gaussian random variable. This is established in the following lemma.
\begin{lemma}
\label{sum_truncated_gaussian_PDF}
\normalfont
Let $X  \sim \mathcal{N}_A(0,\sigma_x^2)$, $N \sim \mathcal{N}(0,\sigma^2)$ and $T=X+N$. Then the PDF of $T$, $p_T(t)$, is given by

\vspace{-.4cm}
\begin{small}
\begin{align}
\label{sum_truncated_gaussian_gaussian_PDF}
p_T(t)&= g(t) w(t),
\end{align}
\end{small}

\vspace{-.4cm}
where $g(\cdot)$ and $w(\cdot)$ are given by:

\vspace{-.4cm}
\begin{small}
\begin{align}
\label{equivalent_normal}
g(t)&=\frac{1}{\sqrt{2 \pi (\sigma^2+\sigma_x^2)}}  \exp(-\frac{t^2}{2  (\sigma ^2+ \sigma_x ^2)}) \\
\label{weighing_function}
w(t)&=\frac{\mathcal{Q}(\frac{-A-t\frac{\tilde{\sigma}^2}{\sigma^2}}{\tilde{\sigma}}) - \mathcal{Q}(\frac{A-t\frac{\tilde{\sigma}^2}{\sigma^2}}{\tilde{\sigma}})}{D}.
\end{align}
\end{small}
\end{lemma}
\begin{IEEEproof}
See Appendix.
\end{IEEEproof}
An analytic expression for the optimal input variance $\sigma_x^2$ is difficult to obtain. However, based on numerical observations, $\sigma_x^2$  can be set to a simple heuristic value given by $\hat \sigma_x^2=A^2$.
\subsection{Lower and Upper bounds}
Based on \eref{equivalent_wiretap}, the secret-key capacity can be lower bounded as follows

\vspace{-.4cm}
\begin{small}
\begin{align}
C_k&=  \max_{p_X(x)  } \left[ \mathbb{I}(X,X+N_{eq})-\mathbb{I}(X;Z)\right] \nonumber\\
& \geq   \max_{p_X(x)  }  \mathbb{I}(X,X+N_{eq}) -    \max_{p_X(x)  }  \mathbb{I}(X;Z)  \\
&= C_{BE}-C_{E}  \triangleq C_{k,1}^{LB},
\label{lower_bound_1}
\end{align}
\end{small}

\vspace{-.4cm}
where $C_{BE}$ corresponds to the capacity of the \textit{enhanced} receiver channel  with input $X$ and output$ \frac{Y}{\sigma_D^2}+\frac{Z}{\sigma_E^2}$ and $C_{E}$ corresponds to the capacity of the eavesdropper channel. 
Having closed-form expressions is of interest in many scenarios. Next, closed-form lower bounds are provided. Considering \eref{equivalent_wiretap} and using 
\cite[Theorem 1]{MISO_visible_light}, the following lower bounds follow immediately

\vspace{-.4cm}
\begin{footnotesize}
\begin{align}
\label{lower_bound_2}
C_k& \geq \frac{1}{2} \log (1+ \frac{ \frac{2A^2}{\sigma_{DE}^2} }{\pi e})-(1-2 \mathcal{Q}(\beta+\frac{A}{\sigma_E})) \log (  \frac{2 (\beta +\frac{A}{\sigma_E})}{\sqrt{2 \pi} (1-2 \mathcal{Q}(\beta))}) \nonumber\\
&
-\mathcal{Q}(\beta)-\frac{\beta}{\sqrt{2 \pi}} e^{-\frac{\beta^2}{2}}+\frac{1}{2}\triangleq  C_{k,2}^{LB}
\\
C_k& \geq \frac{1}{2} \log \frac{6 \frac{A^2}{\sigma_{DE}^2}+3 \pi e}{\pi e \frac{A^2}{\sigma_{E}^2}+3 \pi e} \triangleq  C_{k,3}^{LB},
\label{lower_bound_3}
\end{align}
\end{footnotesize}

\vspace{-.4cm}
where $\beta >0$ is a free parameter. Since \eref{lower_bound_2} holds true for any $\beta>0$, one may maximize $C_{k,2}^{LB}$ over all $\beta$.  Following \cite[Theorem 1]{MISO_visible_light}, \eref{lower_bound_2} and \eref{lower_bound_3} can be derived by using the dual expression for the secrecy capacity and choosing a particular distribution that results in a secrecy achievable rate.  

Clearly, The secret-key capacity with an amplitude constraint $A$ is upper bounded by the secret-key capacity with an average power constraint $A^2$. It follows that

\vspace{-.4cm}
\begin{small}
\begin{align}
\label{Upper_bound}
C_k& \le \frac{1}{2} \log( 1+ \frac{A^2\sigma_E^2}{(A^2+\sigma_E^2)\sigma_D^2}  )\triangleq C_{k}^{UB}.
\end{align}
\end{small}
\vspace{-.6cm}
\subsection{Asymptotic analysis}
First, the secret-key capacity is investigated for asymptotically high values of A.
\begin{proposition}
\normalfont
\label{asymptotic_capacity}
The secret-key capacity with amplitude constraint coincides with the secret-key capacity with an average power constraint and converges to $\frac{1}{2} \log (1+\frac{\sigma_E^2}{\sigma_D^2} )$ as $A$ goes to infinity. 
\end{proposition}
\begin{IEEEproof}
Setting $\beta= \log(1+\frac{2A}{\sigma_E})
$ as in \cite{MISO_visible_light}, it can be seen that the upper bound $C_{k}^{UB}$ and the lower bound $C_{k,2}^{LB}$ converge to the same value for asymptotically high values of $A$, given by $\frac{1}{2} \log (1+\frac{\sigma_E^2}{\sigma_D^2} )$. Thus, one obtains

\vspace{-.4cm}
\begin{small}
\begin{align*}
\lim_{A \to \infty} C_k=\lim_{A \to \infty} C_{k,2}^{LB}(\beta^*)=\lim_{A \to \infty} C_k^{UB}=\frac{1}{2} \log (1+\frac{\sigma_E^2}{\sigma_D^2}).
\end{align*}
\end{small}
\end{IEEEproof}
When $A$ is very small, the behavior of the secret-key capacity is similar to the capacity in the absence of an eavesdropper.
\begin{proposition}
\normalfont
\label{low_SNR_regime}
In the regime $\frac{A}{\sigma_{DE}} <<1$, the secret-key capacity behaves as the capacity of the legitimate user channel with no secrecy constraint as follows

\vspace{-.1cm}
\begin{small}
\begin{align}
\label{lower_bound_behavior}
\lim_{\frac{A}{\sigma_{DE}} \to 0} \frac{C_k}{\frac{1}{2} \frac{A^2}{\sigma_{D}^2}}=1.
\end{align}
\end{small}

\vspace{-.4cm}
\end{proposition}
\begin{IEEEproof}
First, the asymptotic equivalent expression of the capacity for a Gaussian channel with no secrecy constraint is established. The result is summarized in the following lemma.
\begin{lemma}
\label{lemma_low_A_classic}
\normalfont
Consider a Gaussian channel with amplitude constraint $A$ with capacity $C$ given by $Y=X+N$ such that $N \sim \mathcal{N}(0,\sigma^2)$. Then, it follows that 
\begin{align}
\lim_{ \frac{A}{\sigma} \to 0} \frac{C}{ \frac{A^2}{2 \sigma^2} }=1.
\end{align} 
\end{lemma}
\begin{IEEEproof}
Without loss of generality, assume $\sigma=1$ (in which case $A$ represents is the signal amplitude to noise standard deviation ratio). A lower bound of $C$ is obtained by considering an equal pair of mass points at the interval extremes as a particular input distribution. The rate in this case is
\begin{align}
R=h(Y) -\frac{1}{2} \log(2 \pi e),
\end{align}
where $h(Y)$ corresponds to the entropy of a mixed Gaussian distribution.
From \cite{michalowicz2008calculation}, $h(Y)$ is given by 
\begin{align}
&h(Y)=\frac{1}{2} \log (2 \pi e )+A^2-I \nonumber \\
&I= \frac{2}{\sqrt{2 \pi}  A} \exp(-\frac{A^2}{2}) \\
& \quad \times \int_{0}^{\infty} \exp(-\frac{y^2}{2 A^2}) \cosh(y) \log(\cosh(y)) dy
\end{align}
So that one obtains
\begin{align}
\label{rate_expression}
R=A^2-I.
\end{align}
An upper bound of $C $ is the capacity with average power constraint $A^2$. On the other hand, $ R \geq \frac{A^2}{2}-I$. Thus, one obtains
\begin{align}
\label{equivalence}
\frac{A^2}{2}-I \le  C \le \frac{1}{2} \log(1+ A^2)
\end{align}
Clearly,  
\begin{align}
\label{upper_bound_equivalence}
\lim_{A \to 0} \frac{ \frac{1}{2} \log(1+\frac{A^2}{2})}{   \frac{A^2}{2} } =1.
\end{align}
Now, it remains to prove that
\begin{align}
\lim_{A \to 0 } \frac{I}{\frac{A^2}{2}}=0 \iff \lim_{A \to 0 } \frac{I}{A^2}=0.
\end{align}

For that, one writes

\begin{footnotesize}
\begin{align}
\frac{I}{A^2}&=  \frac{2}{\sqrt{2 \pi} }  \int_{0}^{\infty} 
\frac{1}{A^3}   \exp(-\frac{A^2}{2}) \exp(-\frac{y^2}{2 A^2})  
\cosh(y) \log(\cosh(y)) dy.\\
&= \mathds{E}_T \left[  \frac{2}{A^2}  \log \cosh (A T )]  \right],
\end{align}
\end{footnotesize}

with $T$ is a mixed Gaussian distribution $ \sim f_T(t)=\frac{1}{\sqrt{2 \pi}} \exp(-\frac{t^2+A^2}{2}) \cosh(A t)$.
Using the inequality $ \cosh(x) \le \exp(\frac{x^2}{2})$, we have,
\begin{align}
\label{upper_limit_BCT}
\mathds{E}_T  [  \frac{2}{A^2}  \log \cosh (A T )]   ] \le 
\mathds{E}_T  [  T^2    ]=1+A^2.
\end{align}
Note that the right-hand side term of \eref{upper_limit_BCT} is bounded as $A \to 0$. Similarly, using the inequality $\exp(x) \le \cosh(x)$, we write
\begin{align}
\label{lower_limit_BCT}
\mathds{E}_T  [  \frac{2}{A^2}  \log \cosh (A T )]   ] \geq
\mathds{E}_T  [   \frac{2}{A }  T    ]=0.
\end{align}
On the other hand,  
\begin{align}
\label{inside_limit}
\forall y >0, \lim_{A \to 0} \frac{1}{A^3}   \exp(-\frac{A^2}{2}) \exp(-\frac{y^2}{2 A^2}) =0
\end{align}
Then, combining \eref{upper_limit_BCT}, \eref{lower_limit_BCT}, \eref{inside_limit} and applying the dominated convergence theorem, it follows
\begin{align}
\lim_{A \to 0 } \frac{I}{A^2}=0,
\end{align}
which implies 
\begin{align}
\lim_{A \to 0} \frac{\frac{A^2}{2}-I}{\frac{A^2}{2}}=1.
\label{lower_bound_equivalence}
\end{align}
Therefore, combining \eref{equivalence}, \eref{upper_bound_equivalence} and \eref{lower_bound_equivalence}, one obtains
\begin{align}
\lim_{A \to 0} \frac{C}{\frac{A^2}{2}}=1.
\end{align}
\end{IEEEproof}
Going back to the proof of \pref{low_SNR_regime} and assuming that $\frac{A}{\sigma_{DE}} << 1$, it follows that $\frac{A}{\sigma_{E}} << 1$ and $ \frac{A}{\sigma_{D}} << 1$. Applying the result of \lref{lemma_low_A_classic} to \eref{lower_bound_1}, one obtains
\begin{align}
\lim_{ \frac{A}{\sigma_{DE}} \to 0} \frac{C_{k,1}^{LB}}{ \frac{A^2}{2 \sigma_D^2} }
&=  
\frac{\sigma_D^2}{\sigma_{DE}^2}  \lim_{ \frac{A}{\sigma_{DE}} \to 0}( \frac{C_{BE}}{ \frac{A^2}{2 \sigma_{DE}^2} })
- 
\frac{\sigma_D^2}{\sigma_{ E}^2}  \lim_{ \frac{A}{\sigma_{DE}} \to 0}( \frac{C_{ E}}{ \frac{A^2}{2 \sigma_E^2} }).
\nonumber\\
&=\frac{\sigma_D^2}{\sigma_{DE}^2}-\frac{\sigma_D^2}{\sigma_{ E}^2}=1.
\label{low_bound_behavior}
\end{align}
Moreover, from \eref{Upper_bound}, one obtains

\vspace{-.4cm} 
\begin{small}
\begin{align}
\lim_{\frac{A}{\sigma_{DE}} \to 0} \frac{C_k^{UB}}{\frac{A^2}{2 \sigma_D^2}}=
\lim_{\frac{A}{\sigma_{DE}} \to 0} \frac{\log(1+ \frac{A^2}{\sigma_D^2 (1+\frac{A^2}{\sigma_E^2})}) }{\frac{A^2}{ \sigma_D^2}}=1.
\label{upp_bound_behavior}
\end{align}
\end{small}

\vspace{-.4cm}

The result follows by combining \eref{low_bound_behavior} and \eref{upp_bound_behavior}.
\end{IEEEproof}
\section{Numerical results}\label{S5}
\begin{figure}
\centering
\includegraphics[width=.9\linewidth]{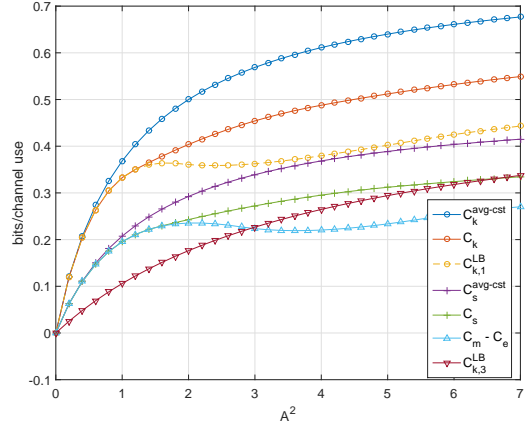}
\caption{The secret-key and secrecy capacities versus the square of the peak amplitude constraint A.}
\label{fig:capacities_vs_A2}
\end{figure}
In this section, numerical results are presented to validate the analytical derivations. Figure 1 illustrates numerically the result in Theorem 1 by computing the secret-key capacity for $\sigma_D^2=1$ and $\sigma_E^2=2$. 
The secret-key capacity with an average power constraint $A^2$, denoted $C_k^{UB}$, is also visualized. 
For comparison, Fig. 1 illustrates the secrecy capacity under an amplitude constraint $C_{s}$, the secrecy capacity under an average power constraint $C_{s}^{avg-cst}$, the lower bounds $C_{k,1}^{LB}$ and $C_{k,3}^{LB}$ ($C_{k,2}^{LB}$ is negative over the considered range) and a lower bound to the secrecy capacity $C_m-C_e$, with $C_m$ being the capacity of the legitimate receiver. It is observed that the rate of increase of $C_k$ and $C_k^{avg-cst}$ are the same, which parallels the observations for the capacity with no secrecy constraint \cite{smith1971information} and the secrecy capacity \cite{ozel2015gaussian}. 
For low values of $A$, one observes that the secret-key capacity with an amplitude constraint coincides with its lower bound $C_{k,1}^{LB}$ and is very close the secret-key capacity with an average power constraint. 
It is worth noting that as $A$ grows, the optimal number of mass points, as well as the optimization complexity, increases.
\\
Next, the performance of the proposed suboptimal schemes is illustrated in Fig. 2. The noise variances are set to $\sigma_D=1$ and $\sigma_E=1.5$. The maxentropic distribution coincides with the secret-key capacity for low values of $A$ and achieves near capacity rates for higher values of $A$. Moreover, the heuristic choice of $K$ behaves very close to the optimized maxentropic distribution. On the other hand, the continuous uniform distribution and the truncated Gaussian perform less efficiently for low values of $A$. However, for high values of $A$, the truncated Gaussian distribution (as well as the one with the heuristic choice of $\sigma_x=A$) outperforms all other suboptimal schemes and achieves rates very close to the capacity. 
\\
Finally, Fig. 3 illustrates numerically the convergence of $C_{k,2}^{LB}(\beta^*)$ and $C_k^{UB}$ to the same limit as $A$ goes to infinity, which confirms the result in \pref{asymptotic_capacity}.
\begin{figure}
\centering
\includegraphics[width=.9\linewidth]{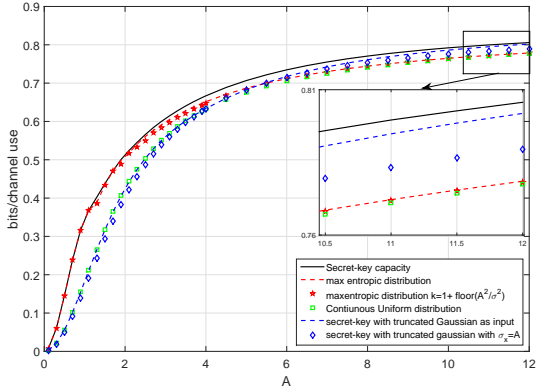}
\caption{Secret-key rates for different schemes.}
\label{fig:different_schemes}
\end{figure} 

\begin{figure}
\centering
\includegraphics[width=.9\linewidth]{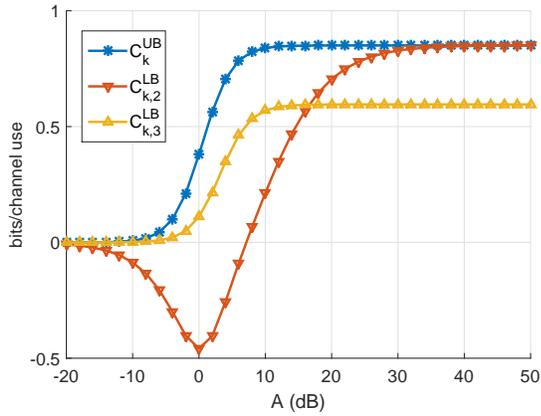}
\caption{The secret-key and secrecy capacities versus the square of the peak amplitude constraint A.}
\label{fig:bounds}
\end{figure}
\section{Conclusion}\label{S6}
In this paper, the capacity-achieving probability distribution of the channel model for secret-key agreement is studied. The optimal distribution is proved to discrete with finite number of probability mass points. To circumvent the numerical complexity associated with finding the optimal input distribution, several schemes are proposed and shown numerically that they perform close to the capacity. Moreover, upper and lower bounds to the secret-key capacity are provided and used to prove that that the secret-key capacity under amplitude constraint $A$ coincides with the capacity under an average power constraint $A^2$ for asymptotically high values of $A$. Finally, it is shown that, when $A$ is small, the secret-key capacity behaves as the capacity with an amplitude constraint A and no secrecy constraint.
\section*{Appendix}
\subsection{Proof of \lref{sum_truncated_gaussian_PDF}}
\label{appendix_proof_sum_gaussian_trunc_PDF}

\vspace{-0.4cm}
\begin{small}
\begin{align}
\label{pdf_eq0}
& f(y)\nonumber\\
&= \int_{-A}^A f(y|x) dF(x) \\
\label{pdf_eq1}
&= \frac{1}{D\sqrt{2 \pi \sigma_x^2} \sqrt{2 \pi \sigma^2} } \int_{-A}^A \exp(-\frac{x^2}{2 \sigma_x^2}) \exp(- \frac{(x-y)^2}{2\sigma^2} ) dx \\
\label{pdf_eq2}
&=\frac{1}{D\sqrt{2 \pi \sigma_x^2} \sqrt{2 \pi \sigma^2} } \exp(-\frac{y^2}{2 \sigma^2 }) 
\int_{-A}^A \exp(-\frac{x^2}{2 \tilde{\sigma}^2 }) + \frac{xy}{\sigma^2} dx\\
\label{pdf_eq3}
&=\frac{\exp(-\frac{y^2}{2  \sigma ^2 } (1-\frac{\tilde{\sigma}^2}{\sigma^2})) }{D\sqrt{2 \pi \sigma_x^2} \sqrt{2 \pi \sigma^2} }
\int_{-A}^A \exp(-\frac{ ( x- y \frac{\tilde{\sigma}^2}{\sigma^2} )^2}{2 \tilde{\sigma}^2 } ) dx\\
\label{pdf_eq4}
&=  \frac{ \exp(-\frac{y^2}{2  (\sigma ^2+ \sigma_x ^2)})}{\sqrt{2 \pi (\sigma^2+\sigma_x^2)}} 
\frac{1}{D} \frac{1}{\sqrt{2 \pi \tilde{\sigma}^2}}  \int_{-A}^A \exp(-\frac{ ( x- y \frac{\tilde{\sigma}^2}{\sigma^2} )^2}{2 \tilde{\sigma}^2 } ) dx \\
\label{pdf_eq5}
&=\frac{\exp(-\frac{y^2}{2  (\sigma ^2+ \sigma_x ^2)})}{\sqrt{2 \pi (\sigma^2+\sigma_x^2)}}  
\frac{1}{D} P ( \mathcal{N}( y \frac{\tilde{\sigma}^2}{\sigma^2}, \tilde{\sigma}^2) \in [-A, A ])\\
\label{pdf_eq6}
&=\underbrace{
\frac{\exp(-\frac{y^2}{2  (\sigma ^2+ \sigma_x ^2)})}{\sqrt{2 \pi (\sigma^2+\sigma_x^2)}}   }_{g(y)}  \underbrace{
\frac{1}{D} P ( \mathcal{N}(0,1)  \in [\frac{-A-y\frac{\tilde{\sigma}^2}{\sigma^2}}{\tilde{\sigma}}, \frac{A-y\frac{\tilde{\sigma}^2}{\sigma^2}}{\tilde{\sigma}} ])}_{w(y)},
\end{align}
\end{small}

\vspace{-0.4cm}
where  \eref{pdf_eq2} is obtained by setting $\frac{1}{\tilde{\sigma}^2} \triangleq \frac{1}{\sigma_x^2}+ \frac{1}{\sigma ^2}$. Equation \eref{pdf_eq3} follows form the fact that $\frac{ 1-\frac{\tilde{\sigma}^2}{\sigma^2} }{   \sigma ^2 }  =\frac{1}{\sigma^2+\sigma_x^2}$, multiplying the numerator and the denumerator by $ \sqrt{2 \pi \tilde{\sigma}^2}$ and the identity $\frac{ \tilde{\sigma}^2 }{  \sigma_x ^2 \sigma ^2 } =\frac{1}{\sigma^2+\sigma_x^2}$. Finally, note that $w(y)$ can be expressed in terms of the $\mathcal{Q}$ function as in \eref{weighing_function}.
%
\bibliography{biblio}

\begin{thebibliography}{10}
\providecommand{\url}[1]{#1}
\csname url@samestyle\endcsname
\providecommand{\newblock}{\relax}
\providecommand{\bibinfo}[2]{#2}
\providecommand{\BIBentrySTDinterwordspacing}{\spaceskip=0pt\relax}
\providecommand{\BIBentryALTinterwordstretchfactor}{4}
\providecommand{\BIBentryALTinterwordspacing}{\spaceskip=\fontdimen2\font plus
\BIBentryALTinterwordstretchfactor\fontdimen3\font minus
  \fontdimen4\font\relax}
\providecommand{\BIBforeignlanguage}[2]{{%
\expandafter\ifx\csname l@#1\endcsname\relax
\typeout{** WARNING: IEEEtran.bst: No hyphenation pattern has been}%
\typeout{** loaded for the language `#1'. Using the pattern for}%
\typeout{** the default language instead.}%
\else
\language=\csname l@#1\endcsname
\fi
#2}}
\providecommand{\BIBdecl}{\relax}
\BIBdecl

\bibitem{common_randomness}
R.~Ahlswede and I.~Csiszar, ``Common randomness in information theory and
  cryptography. i. secret sharing,'' \emph{IEEE Trans. Inf. Theory}, vol.~39,
  no.~4, pp. 1121--1132, Jul 1993.

\bibitem{wongsecret2009}
T.~F. Wong, M.~Bloch, and J.~M. Shea, ``Secret sharing over fast-fading {MIMO}
  wiretap channels,'' \emph{EURASIP J WIREL COMM}, vol. 2009, p.~8, 2009.

\bibitem{Maurer}
U.~Maurer, ``Secret key agreement by public discussion from common
  information,'' \emph{IEEE Trans. Inf. Theory}, vol.~39, no.~3, pp. 733--742,
  May 1993.

\bibitem{smith1971information}
J.~G. Smith, ``The information capacity of amplitude-and variance-constrained
  sclar {G}aussian channels,'' \emph{Information and Control}, vol.~18, no.~3,
  pp. 203--219, 1971.

\bibitem{chan2005capacity}
T.~H. Chan, S.~Hranilovic, and F.~R. Kschischang, ``Capacity-achieving
  probability measure for conditionally {G}aussian channels with bounded
  inputs,'' \emph{IEEE Trans. Inf. Theory}, vol.~51, no.~6, pp. 2073--2088,
  2005.

\bibitem{discreteness_study}
A.~Tchamkerten, ``On the discreteness of capacity-achieving distributions,''
  \emph{IEEE Trans. Inf. Theory}, vol.~50, no.~11, pp. 2773--2778, Nov 2004.

\bibitem{perf_analysis}
H.~Lei, H.~Zhang, I.~Ansari, C.~Gao, Y.~Guo, G.~Pan, and K.~Qaraqe,
  ``Performance analysis of physical layer security over generalized- k fading
  channels using a mixture gamma distribution,'' \emph{IEEE Commun. Lett},
  vol.~20, no.~2, pp. 408--411, Feb 2016.

\bibitem{agrawal2011noncoherent}
A.~Agrawal, Z.~Rezki, A.~J. Khisti, and M.-S. Alouini, ``Noncoherent capacity
  of secret-key agreement with public discussion,'' \emph{IEEE Trans. Inf.
  Forensics Security}, vol.~6, no.~3, pp. 565--574, 2011.

\bibitem{ozel2015gaussian}
O.~Ozel, E.~Ekrem, and S.~Ulukus, ``{G}aussian wiretap channel with amplitude
  and variance constraints,'' \emph{IEEE Trans. Inf. Theory}, vol.~61, no.~10,
  pp. 5553--5563, 2015.

\bibitem{max_entropic}
A.~Farid and S.~Hranilovic, ``Channel capacity and non-uniform signalling for
  free-space optical intensity channels,'' \emph{IEEE J. Sel. Areas Commun},
  vol.~27, no.~9, pp. 1553--1563, December 2009.

\bibitem{conrad2013probability}
\BIBentryALTinterwordspacing
K.~Conrad, ``Probability distributions and maximum entropy.'' [Online].
  Available:
  \url{http://www.math.uconn.edu/~kconrad/blurbs/analysis/entropypost.pdf}
\BIBentrySTDinterwordspacing

\bibitem{truncated_gauss}
A.~Chaaban, Z.~Rezki, and M.-S. Alouini, ``On the capacity of the
  intensity-modulation direct-detection optical broadcast channel,'' \emph{IEEE
  Trans. Wireless Commun}, vol.~PP, no.~99, pp. 1--1, 2016.

\bibitem{MISO_visible_light}
A.~Mostafa and L.~Lampe, ``Physical-layer security for {MISO} visible light
  communication channels,'' \emph{IEEE J. Sel. Areas Commun}, vol.~PP, no.~99,
  pp. 1--1, 2015.

\bibitem{michalowicz2008calculation}
J.~V. Michalowicz, J.~M. Nichols, and F.~Bucholtz, ``Calculation of
  differential entropy for a mixed gaussian distribution,'' \emph{Entropy},
  vol.~10, no.~3, pp. 200--206, 2008.

\end{thebibliography}
\bibliographystyle{IEEEtran}

\end{document}